\documentclass[oneside,10pt]{article}
\usepackage{epsfig}
\usepackage{graphicx}
\usepackage{epstopdf}
\def\ni{\noindent}
\newcommand{\be}{\begin{equation}}
\newcommand{\ee}{\end{equation} \noindent}

\begin{document}
\pagestyle{empty}

\begin{flushright}
 17 Mar 2009
\end{flushright}

 \begin{center} 
\large
Alternative cosmology fits supernovae redshifts with no dark energy

Francis.J.M. Farley\footnote{\emph{email: F.Farley @ soton.ac.uk}} \\
\small \emph{School of Engineering and the Environment \\ Southampton University, Highfield, Southampton SO17 1BJ, England}
\end{center}


\small
Supernovae and radio galaxy redshift data are fitted in an alternative cosmology.  The galaxies are assumed to recede with unchanging velocities in a static Robertson-Walker metric with $a(t)=1$.  An exact fit is obtained with no adjustable parameters.  There is no indication that the recession velocities are changing with time, so no call for ''dark energy''.



 \normalsize 
 
Many studies of supernovae of type 1a (SN1a) have been reported.  It is established that with few exceptions their intrinsic luminosity is the same, so they can be used as standard candles.  Their apparent magnitude $M$ is then a measure of their distance from earth when the light was emitted.  The redshift $z$ indicates the velocity of recession.  The  Hubble plot of $M$ vs $z$ gives information about the recession of the galaxies.  Corresponding data are available on some radio galaxies.
A recent compilation of the data on SN1a, with many references, has been given by Kowalski et al \cite{kow} and further data has been added by Hicken et al \cite{hic}.  

The standard interpretation of the redshift data \cite{riess, perl, cook} uses a flat Robertson-Walker metric
\be
(ds)^2 = -(dt)^2 + a(t) \{(dx)^2 + (dy)^2 + (dz)^2\}  \label{metric}
\ee
The galaxies are stationary in their local space, but space is expanding because $a(t)$ increases with time.  There was no redshift when the light was emitted, but as space has expanded while the photon was traveling, its wavelength has been stretched to correspond.  The theory has been extensively reviewed and compared with observations \cite{cop,perl2,frieman}.  Two parameters $\Omega_M$ and $\Omega_\Lambda$ are adjusted but this is insufficient to fit the data.  It is concluded that the expansion must be speeding up at a rate determined by a third parameter, which can be adjusted to give a good fit.   (This is not enough for some authors \cite{frieman,daly} who can fit any curve by allowing the "cosmic acceleration" to vary with the redshift).  The accelerated expansion is ascribed to ''dark energy'', which remains a mystery \cite{ellis}.

Is there an alternative hypothesis?  Here we examine a model  with no adjustable parameters, which to our surprise fits the SN1a data perfectly.  Perhaps space is not expanding, $a(t)$ is constant, but the galaxies are moving through it at 
 unchanging velocities, diverging from some point at which they (or their constituent primordial matter) were originally close together.  An observer at any point, not necessarily the original centre, will see the other galaxies moving away from him.  The distance reached when the light was emitted will be a function of the unchanging velocity $\beta  c$ of a galaxy relative to the observer.  In this picture the light emitted by a moving supernova is transformed to the rest frame of the observer, assumed stationary, using the Lorentz transformation.  The Doppler shift and solid angle adjustment occur immediately and the photon then propagates unchanged to the receiver.  All calculations use the formulae of special relativity which have been accurately verified at extreme velocities \cite{bailey} . The reported magnitudes of SN1a are taken to represent the energy flux received at earth at the peak of the luminosity curve \cite{peak}.  The luminosity varies as $1/r^2$ where $r$ is the distance from earth when the light was emitted.  In applying the Lorentz transformation of special relativity the following factors must be included:

\ni 1) The redshift $z$ is given by $\lambda = \lambda _0 \ (1 + z)$ with
\be
1 + z = \sqrt { \frac{1+\beta}{1-\beta} }     \label{z}
\ee
where the velocity of recession is $\beta c$.  This implies
\be
\beta = \frac{2z+z^2}{2+2z+z^2}   \label{beta}
\ee
\ni 2)  Transformation of solid angle, Jacobian.  For light emitted in the backward direction by a moving source, the transverse momentum is not affected by the motion while the longitudinal momentum is reduced by the factor $1/(1+z)$.  As a result the solid angle of the emitted light is increased by $(1+z)^2$.  The luminosity is decreased by the factor $1 / (1+z)^2$.

\ni 3) The energy of the photons emitted backwards is reduced by the factor $1 / (1+z)$.  As stellar magnitude is calculated from the energy flux, there is a corresponding reduction in the luminosity.

\ni 4) Time dilation.  For an observer at rest, everything in the frame moving away from him is apparently slowed down: time is dilated by the factor $(1+z)$.  The redshift noted above is an example of this.  It applies also to the temporal evolution of the supernova luminosity, which has been carefully verified by Goldhaber et al {\cite{gold}.  As the received light is spread out over a longer time, the peak of the luminosity curve must be decreased by the factor $1 / (1+z)$.

\begin{figure}
\begin{center}
\resizebox{8cm}{!}{ \includegraphics{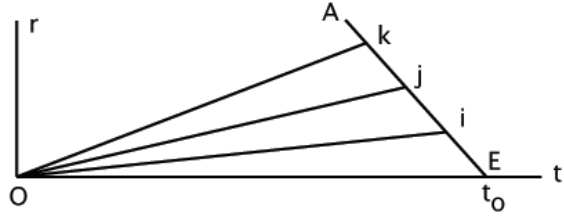}}
\caption{Supernova distance $r$ vs. time}
\label{f1}
\end{center}
\end{figure}

\ni 5) The time for the light to travel from the SN to earth must be considered; distant SN emitted their light long before it is received.  This is illustrated in Figure 1, which shows the distance $r$ of the SN from earth versus the time $t$ from the big bang.  The SN diverge from time zero along lines \emph{Oi, Oj, Ok} at various velocities $\beta c$, while the light received at earth at time $t_0$ (now) follows the line \emph{AE}.  The intersection points \emph{i, j, k} determine the distance $r$ at the moment the light was emitted.  It follows that
\be
r = \frac{\beta c t_0}{1+\beta}  \label{r}
\ee
The received luminosity is proportional to
\be
\frac{1}{r^2} \propto \frac{(1+\beta) ^2}{\beta ^2}  \label{td}
\ee

\ni Combining all these factors and using eqn (\ref{beta}), the observed luminosity $L$ is given by
\be
L = \frac{L_0}{c^2 t_0^2} \frac{(1+\beta) ^2}{\beta ^2}  \frac{1}{(1+z)^4} = \frac{L_0}{c^2 t_0^2}\frac{4}{(2z+z^2)^2}   \label{LU}
\ee
Therefore the observed SN magnitudes at redshift $z$ should be
\be
M = -2.5\  log_{10} (L) = M_0 + 5\  log_{10}(2z+z^2)  \label{M}
\ee
$M_0$ is an arbitrary constant, proportional to the intrinsic luminosity of the standard candle.   The value  $M_0 = 41.8$ is inserted to give the best fit.

The graphs of Figures 2 and 3, at two scales of $z$, show that this prediction corresponds exactly with the ``constitution'' data set recommended by Hicken et al \cite{hic}.  The only adjustable parameter is $M_0$, which moves the graph vertically.  There are no parameters to manipulate the shape of the curve, but one sees that no improvement is required.  Previous authors \cite{dav}, who claim that special relativity does not fit the data, may have omitted one or more of the factors 2--5 listed above.

Data on the apparent magnitudes of 30 radio galaxies versus redshift is presented by Daly et al \cite{daly} and has been added to the graphs of Figures 2 and 3, open squares. This extends the observations to higher redshifts, also in close agreement with equation (\ref{M}).  However the magnitudes tabulated in \cite{daly} are not the raw data, but the result of a complicated fitting process based on general relativity with adjusted parameters; so this agreement with our model is less compelling.

It is surprising that an expression as simple as equation (\ref{M}) with no free parameters can fit SN1a observations so well.  The recession of the galaxies and the supernovae redshifts can apparently be interpreted by this alternative, rather simpler model with the following properties:

\ni 1) The galaxies are receding from each other at unchanging velocities in a static space; \newline
 2) A flat Robertson-Walker metric applies with $a(t)=1$; \newline
 3) The recession velocities are neither speeding up nor slowing down; \newline
 4) Dark energy is not required.

This model may be incompatible with other cosmological data, such as the cosmic microwave background and the distribution of galaxies.  Most phenomena, however, depend on the distribution and movement of matter, not on the expansion of space itself.  These questions remain to be examined.
 
I wish to thank Wim de Boer, Richard Ellis, John Field, Shelly Glashow, Martin Rees, Brian Schmidt, R.K. Thakur and Ivan Zhogin for valuable comments.

\begin{figure}
\begin{center}
\resizebox{16cm}{!}{ \includegraphics{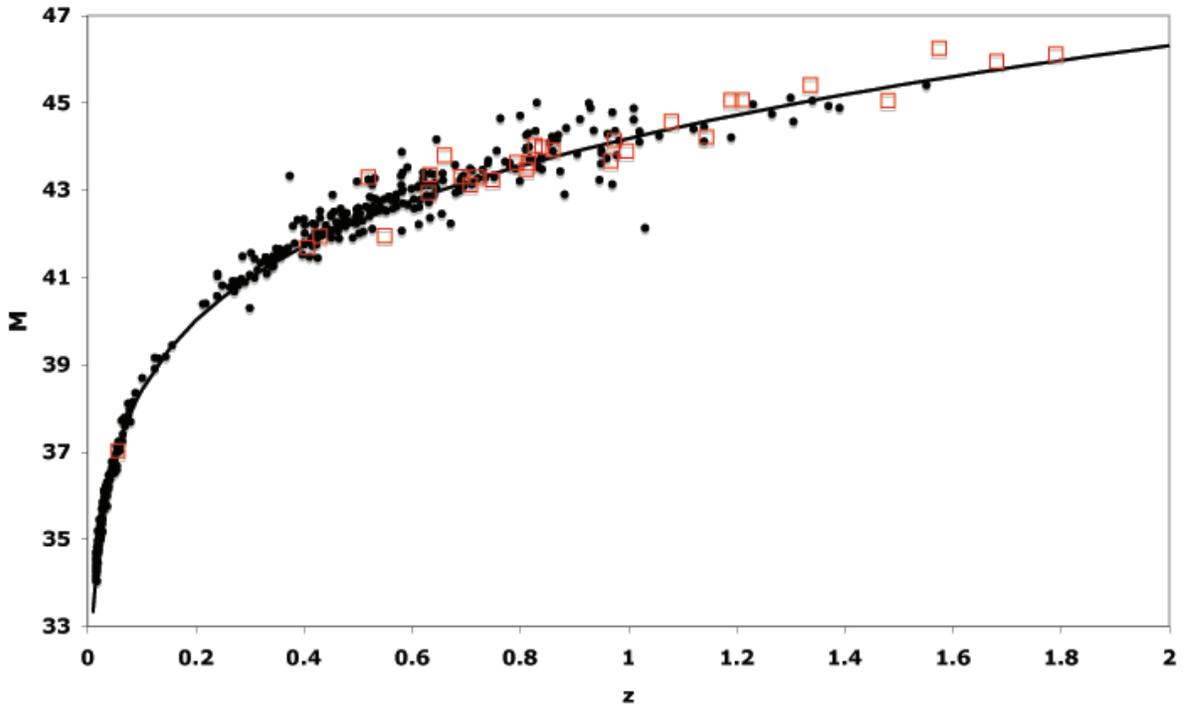}}
\caption{Observed SN1a (black dot) and radio galaxy (open square) magnitudes M vs redshift z compared with equation (7)}
\label{f2}
\end{center}
\end{figure}

\begin{figure}
\begin{center}
\resizebox{16cm}{!}{ \includegraphics{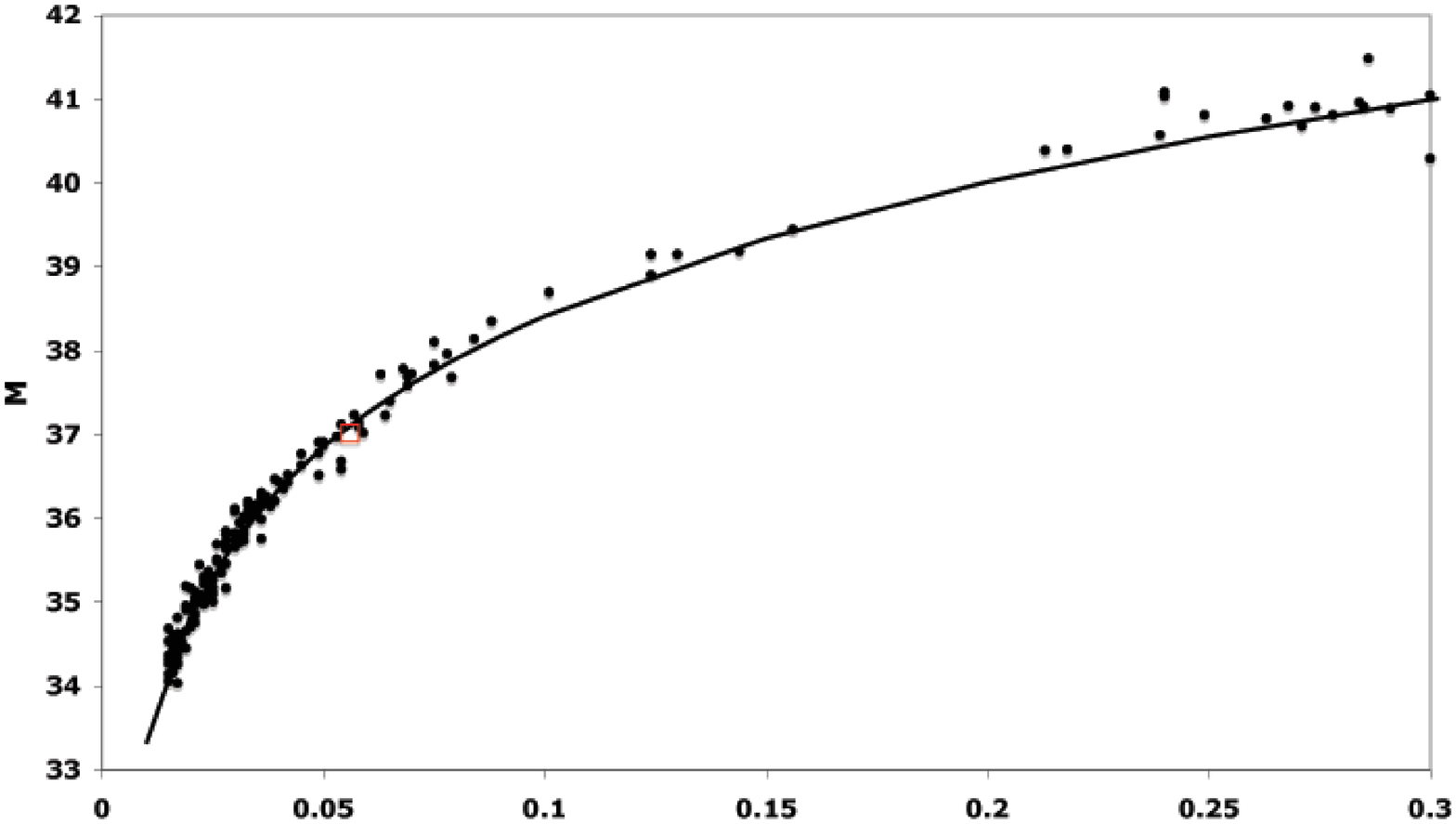}}
\caption{Observed SN1a (black dot) and radio galaxy (open square) magnitudes M vs redshift z compared with equation (7) ... expanded scale}
\label{f3}
\end{center}
\end{figure}

\end{document}